\documentclass[10pt]{iopart}
\usepackage{iopams}  
\usepackage{graphicx}  
\begin{document}

\title[K$_\alpha$ Profiles of Warm Dense Ar]{K$_\alpha$ Emission Profiles of Warm Dense Argon Plasmas}

\author{A Sengebusch$^1$, H Reinholz $^{1,2}$ and G R\"opke$^1$}
\address{$^1$Institute of Physics, University of Rostock, 18051 Rostock, Germany}
\address{$^2$School of Physics, The University of Western Australia, WA 6009 Crawley, Australia}
\ead{andrea.sengebusch@uni-rostock.de}

\vspace{10pt}
\begin{indented}
\item[]July 2017
\end{indented}

\begin{abstract}
K-line profiles emitted from a warm dense plasma environment are used for diagnostics of Ar droplet plasmas created by high energy laser pulses \cite{Ar}. 
Analyzing the temporally and spacially integrated spectra, we infer temperature gradients within the Ar droplet from cold temperatures of the order of some 10 eV up to higher temperatures of 250 eV and beyond.
To characterize the influence of the warm and dense environment on the emitters, plasma screening is considered within a perturbative approach to the Hamiltonian. 
The plasma affects the many-particle system resulting in energy shifts of emission as well as ionization energies due to electron-ion and electron-electron interaction. 
With this approach we get a good reproduction of spectral features that are strongly influenced by ionization and excitation processes within the plasma. Comparing with the widely known FLYCHK code and the temperature distribution given in the original paper \cite{Ar}, counting for internal degrees of freedom (bound states) and treating pressure ionization within our quantum statistical approach results in a more detailed temperature-density-relation and leads to different results for the inferred temperature distribution.
\end{abstract}

\pacs{52.25.Jm, 52.25.Os, 52.38.Ph, 52.50.Jm, 52.65.Vv, 52.70.La}
%
\vspace{2pc}
\begin{indented}
\item Keywords: K-line emission, X-rax spectra, warm dense matter, plasma polarization, plasma composition
\end{indented}
%
\begin{indented}
\item \submitto{\JPB}
\end{indented}
%
\maketitle
%

%


\section{Introduction}

Plasma diagnostics using $K_{\alpha}$ fluorescence spectra allows to investigate properties of warm dense matter. 
This work focuses on the bulk temperature distribution of plasmas created from Ar droplets irradiated by high energy laser pulses 
with a power of 10$^{19}$ W/cm$^2$ \cite{Ar}. Due to the high intensity, matter at temperatures between some 10 eV  to 250 eV (bulk) and up to 1 MeV (blow-off) is created. Most of the atoms are ionized and a wide range of different ion species is observed.

The atomic density of the Ar droplet is $n_{\textrm{\scriptsize tot}}= 2.2 \times 10^{22} $ cm$^{-3}$. 
Accordingly, free electron densities vary between $10^{22}-10^{24}$ cm$^{-3}$ depending on the degree of ionization of the Ar plasma.
Due to laser plasma interaction the argon plasma radiates $K_{\alpha}$ lines with emission energies at about 3 keV, which can be used straightforwardly to infer the plasma parameters.
To propagate through the plasma, radiation frequencies have to be larger than the plasma frequency $\omega_{\textrm{\scriptsize plasma}}=\sqrt{\frac{n_{e}e^{2}}{\epsilon_{0} m}}$, 
where $n_{e}$ is the density and $m$ is the mass of free electrons \cite{saha}. Considering the expected high free electron densities, x-rays (like the $K_{\alpha}$-lines) are needed to investigate the created plasmas.

To describe the plasma microscopically, including its emission spectra, one has to consider in detail the influence of the environment on the ionic potential, the electron-electron and the electron-ion interaction \cite{ion}. To introduce plasma effects, the electron-ion plasma screening is described within an ion sphere model approximation \cite{ion} and the electron-electron plasma screening is described within a quantum statistical approach by the Montroll-Ward self-energy contribution. To understand the $K_{\alpha}$ emissions, which correspond to an electron going from a 2p to an 1s level, also the fine structure of the emitting states has to be taken into account. Moreover, the creation of inner (K-) shell vacancies is accompanied by ionization and excitation of outer shell electrons (L- and M-shell) resulting in additional satellite lines. The emission energies of the satellites significantly increase with the number of vacancies and excited states. The intensity ratio of different lines is determined according to their statistical weight (LS coupling) and the abundance of the corresponding emitter species in dependence on the plasma parameters.
Using coupled Saha equations, we consider the different charge states in thermodynamic equilibrium \cite{saha}. Finally, we add Lorentz profiles and convolute with a Gaussian instrument function to calculate synthetic spectra from the plasma shifted emission energies. 

We compare our results with the experimental data presented in \cite{Ar} as well as with theoretical results obtained with the FLYCHK code, also given in \cite{Ar}. 
FLYCHK\cite{FLYCHKpublication,FLYCHK,FLYCHKManual} provides emission spectra and ionization distributions within plasmas by solving rate equations for level populations considering collisional and radiative atomic processes. The code is well benchmarked for long-pulse laser experiments. However, there can appear jumps in the density of nearly neutral ion species if bound states are pressure ionized. When constructing synthetic X-ray spectra, line components that involve vanishing states have to be removed by hand if there is no steady fade out of its contribution 
with respect to the plasma parameters\cite{Stambulchik09}. We apply an approach where pressure ionized states shift steadily into the continuum and no extra treatment is needed \cite{cpp_chen,Sengebusch09}.

\section{Unperturbed emitter}
To describe the influence of a plasma environment on an emitter, a perturbative ansatz, $H = H_0 + H'$, is chosen. The system’s Hamiltonian $H$ is split into a part $H_0$ describing the isolated emitting particle and a perturbing plasma potential $H'$. The isolated emitter can be described by means of atomic physics. 
In contrast to previous publications, we changed and expanded the emitter configurations that are taken into account. Previously, to determine emission and ionization energies of various ionic configurations, we solved the corresponding self-consistent Roothaan-Hartree-Fock equations applying the chemical ab initio code Gaussian 03, see e.g. \cite{cpp_chen,Sengebusch09}. However the calculations prevent straightforward observation of fine-structure splitting and only a very restricted number of excitation levels could be taken into account. As there are several hundred transitions located closely together, it is common to limit the number of levels in the calculations especially for lower charge states. FLYCHK for example applies super-configuration transition arrays, which result in effective bound states and respective transition rates \cite{Ar}. However, our observations show that the details of the energy level configurations are crucial to the determination of the plasma composition and emission spectra. 
Despite the calculational efforts we decided to take all tabulated values of energy levels and emission energies into account to describe the unperturbed particles as precisely as possible. In particular, we consider ionization for all shells but 1s, i.e. for argon ($Z=18$) we apply 16 ionization stages from Ar$^+$ to Ar$^{16+}$. For those ions we considered altogether 111 different electronic configurations and 713 different ionization energies. The latter can be found in the NIST Atomic Spectra Database \cite{NIST}. 
To get the difference between the two numbers across, let's consider an example: The ground state electronic configuration of Ar$^{+}$ is (1s$^2$2s$^2$2p$^6$3s$^2$3p$^5$). According to LS coupling we have two different states $J=1/2$ and $J=3/2$ with two different ionization energies for the outermost electron. Another possibility is the excited configuration (1s$^2$2s$^2$2p$^6$3s$^2$3p$^4$3d$^1$). Here we already have 28 different ionization energies corresponding to the different terms from $^2$S to $^4$F and J-states from 1/2 to 9/2.
To calculate the plasma composition also the total binding energies of the considered ions were necessary and taken from NIST as well. Moreover, we took into account a total of 1211 different K$_\alpha$ emission lines, as tabulated by Palmeri et al. \cite{palmeri}, to construct synthetic emission spectra. All those level and emission energies build up the unperturbed basis to which a perturbation according to the plasma is added as described below.

According to first order perturbation theory, the unperturbed wave functions are required to calculate changes of energy levels. If we look at single levels, e.g. in ionization processes, the energy shift is simply given by the averaging over the perturbation $H'$
\begin{equation}
\Delta{E^{(1)}}=\langle\varphi_{val}|H'|\varphi_{val}\rangle~.
\end{equation} 
However, when we consider shifts of emission energies two levels are involved. Then the shift is determined by 
\begin{equation}
\Delta{E^{(1)}}=\langle\varphi_i|H'|\varphi_i\rangle - \langle\varphi_f|H'|\varphi_f\rangle~.
\end{equation} 
Both equations apply the unperturbed wave functions. Whereas $\varphi_{val}$ denotes the wave function of the valence orbital of the electron that is ionized, $\varphi_i$ and $\varphi_f$ denote the initial and final orbitals of the emission transition, respectively.  
The orbital wave functions are separated into radial orbitals $R_{nl}(r)$ and spherical harmonics $Y_{lm}(\theta, \phi)$,
\begin{equation}
\varphi_{nlm}(\vec{r})=R_{nl}(r)\cdot Y_{lm}(\theta, \phi)~.
\end{equation} 
Further, the radial atomic orbitals are expanded as a finite superposition of Slater orbitals. The suitable superposition can be determined self-consistently within an iterative process to minimize the energy of Hamiltonian $H_0$. In the following we use the radial wave functions determined by Bunge \textit{et al}. All parameters and results can be found in detail in \cite{Bunge}.

\section{Plasma polarization}

In the following we describe the determination of the perturbing Hamiltonian $H'$ and calculate shifts of the energy levels caused by the surrounding charges in the plasma environment
within the approach described above. To obtain the distribution of the free plasma electrons around the quasi-static ionic emitters, 
an ion sphere or confined atom model is used \cite{cpp_chen}. The ion sphere contains a nucleus of charge $Z=18$ and the respective number of electrons, so that the system in total is neutral. The electron density inside the sphere is divided into the density of bound and free electrons, $n_b(r)$ and $n_f(r)$ respectively. 
Accordingly, the radial Poisson equation for the potential $\phi(r)$ reads
\begin{equation}
\Delta{\phi(r)}=4\pi e~n_b(r)+4\pi e~n_f(r)-4\pi~Ze~\delta(r).\label{poisson}
\end{equation}
The perturbing potential is given by the difference of the sphere's potential with and without free plasma electrons,
\begin{equation}
H'=-e\left[\phi(r)-\phi(r, n_f=0)\right]~.
\end{equation}
The free electron density is determined self-consistently,
\begin{eqnarray}
n_f(r)&=&\frac{4 n_f(R)}{\sqrt{\pi}}\int_{p_0}^{\infty} \frac{dp~p^2}{\left(2mk_BT\right)^{3/2}}~\nonumber\\
&&\times
\textrm{exp}\left[\left( \frac{-p^2}{2mk_BT}+\frac{e~{\phi(r)}}{k_BT} \right)\right],\label{nfree}
\end{eqnarray}
with the minimum momentum $p_0=\sqrt{2me~\phi(r)}$ and the Wigner-Seitz radius $R =\sqrt[3]{\frac{3~Z_\textrm{\scriptsize ion}}{4\pi n_e}}$ which is the boundary of the ion sphere.
Equation (\ref{nfree}) is obtained from the equation of state of non-degenerate, free electrons in momentum representation. The fraction before the integral represents a normalization factor to assure that the number of free electrons within the sphere equals the ionic charge.

The bound electron density is given by the radial wave functions of the unperturbed emitter, 
\begin{equation}
n_b(r)=\frac{1}{4\pi}\sum_{nl}r^2\left|R_{nl}(r)\right|^2.\label{nbound}
\end{equation}
The sum $\sum_{nl}$ depends on the specific electronic configuration of the considered ion (every electron contributes one bound state orbital). Since the configurations of initial and final state of a transition differ, we also obtain different potentials $\phi(r)$ and $H'$ of the ion sphere before and after the transition. For the $R_{nl}(r)$ functions we take the results of Bunge \textit{et al.} \cite{Bunge}, as described above. 

The iterative calculation of $\phi(r)$ and $n_f(r)$ starts with the assumption of a constant free electron density $n_{\textrm{\scriptsize f}}(r)=n_e$ throughout the sphere. 
We solve (\ref{poisson}) and (\ref{nfree}) iteratively and finally obtain a self-consistent free electron density which forms a polarization cloud close to the nucleus and dilutes for larger radii. 

\begin{figure}
\includegraphics[width=0.4\textwidth]{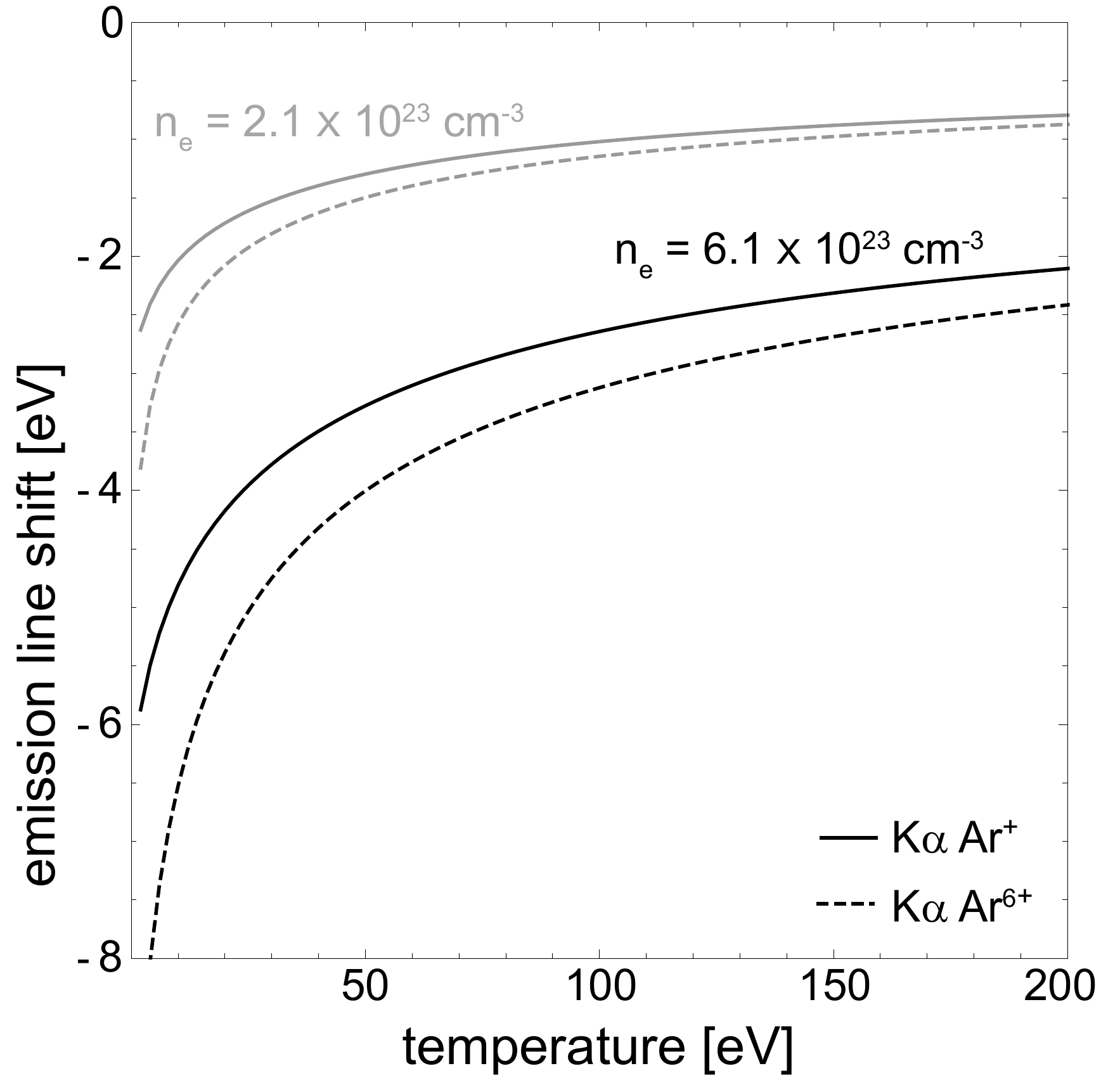}
\caption{Shift of $K_{\alpha}$ emission energy with rising temperature for two different electron densities and ground state ionization stages. }\label{shift_n}
\end{figure}

\begin{figure}
\includegraphics[width=0.4\textwidth]{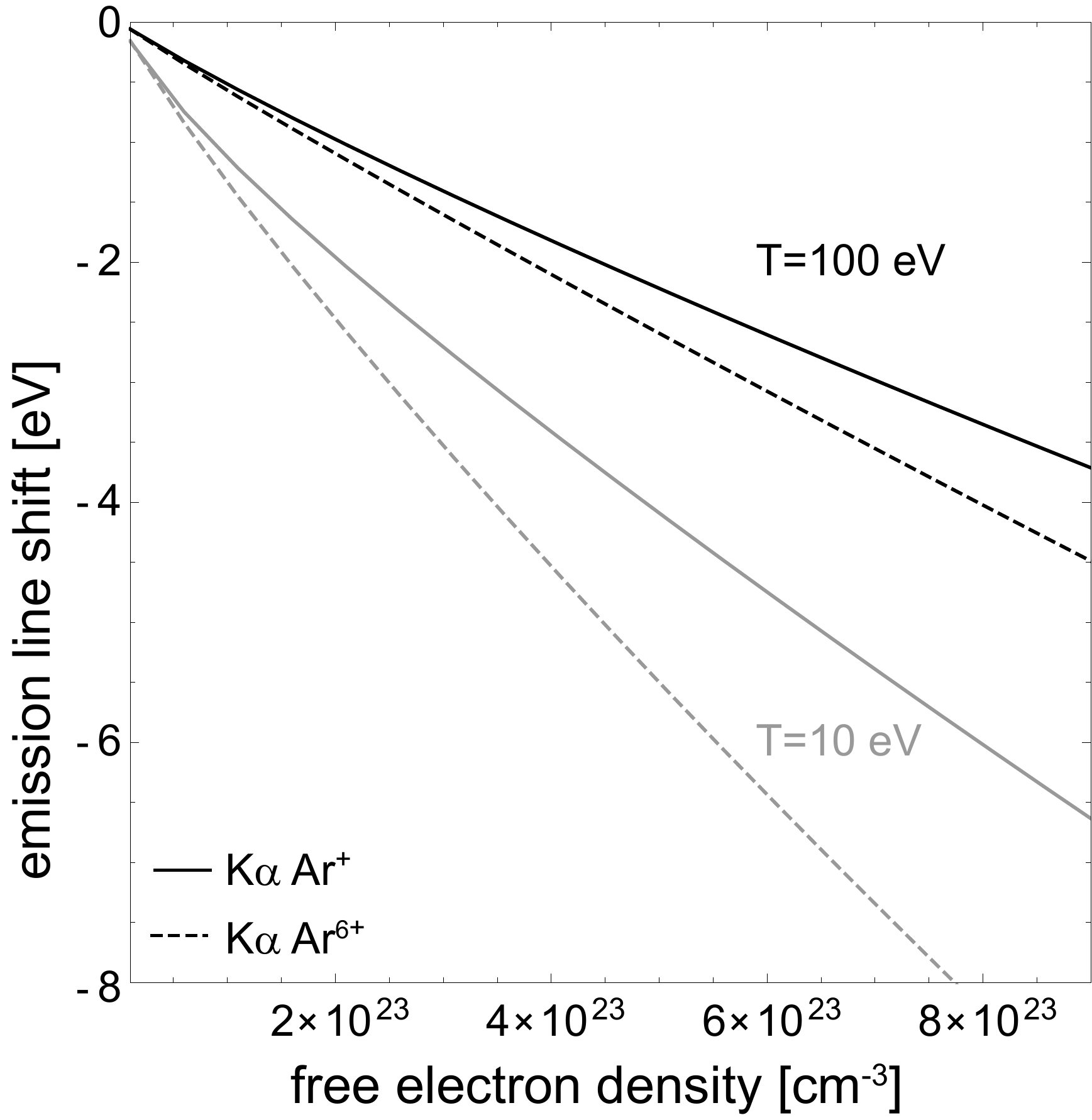}
\caption{Shift of $K_{\alpha}$ emission energy with rising free electron density for two different plasma temperatures and ground state ionization stages.}\label{shift_T}
\end{figure}

Due to screening caused by the free electrons close to the nucleus, the energy levels and hence the emission and ionization energies are shifted. 
Our calculational results are exemplarily shown in figures (\ref{shift_n}) and (\ref{shift_T}). We obtain emission line shifts to lower energies (red shift) in the order of some eV depending on both the plasma temperature $k_BT$ and the average free electron density $n_e$. This effect is referred to as plasma polarization shift and can be explained as follows \cite{Sengebusch09}: 
The free electrons screen the nucleus resulting in lower absolute values of the negative binding energies. As the 1s level is localized closer to the nucleus than 2p, it is more affected by the screening of the nucleus and thus experiences the larger shift, i.e. the gap between the two levels narrows. Since the emission energy is given by the difference of the two involved levels the spectral line is red shifted to lower emission energies. The red shift increases with rising free electron density as the screening of the nucleus rises as well. However, the red shift decreases with rising plasma temperature. This is due to the fact that the self-consistently determined free electron density within the ion sphere is not constant but depends on the distance to the emitter's core. The higher the temperature the more the screening cloud is spatially extended ('smeared out') and the actual free electron density close to the nucleus decreases, resulting in less screening and a lower red shift. 

Plasma polarization is not only important to calculate the accurate emission line positions but also to determine the plasma composition. In contrast to emission lines, a single level surrounded by plasma experiences a blue shift to higher energies when its absolute value of the binding energy is lowered by screening, i.e. the level shifts closer to the continuum ($E=0$). Depending on the plasma parameters this shift can be that large that the continuum is reached and the bound electron is set free (pressure ionization, Mott effect \cite{quantum}). Hence, we obtain another free electron and a new ion species. In the following we will outline, how the plasma composition is calculated taking into account the polarization shifts we determined within the ion sphere model.

\section{Plasma composition}

Assuming a local thermal equilibrium, we apply Saha equations to determine the abundance of the different ion species \cite{CPP_PL},
\begin{equation}
\frac{n_{(m+1)}n_e}{n_{(m)}}=\frac{\sigma_e^{in}}{\lambda_e^3}~\frac{\sigma_{(m+1)}^{in}}{\sigma_{(m)}^{in}}~.
\end{equation}
Here we use a representation of the chemical potential of species $c$ which is of the same form as the one for the classical ideal gas, 
$\mu_c=k_BT~\textrm{ln}\left[\frac{n_c\lambda_c^3}{\sigma_{c}^{in}}\right]$. However, the denominator in the logarithm is given by the internal partition function $\sigma_{c}^{in}$  
instead of the statistical weight of the corresponding particle. The internal partition functions contain the particle interactions and are discussed in detail in the next paragraph. Coupling the Saha equations of successive ionization stages we obtain
\begin{equation}
n_{(m)}=\left[\frac{1}{n_e\lambda_e^3}\right]^m~\sigma_e^{in}(m)~\frac{\sigma_{(m)}^{in}}{\sigma_{(0)}^{in}}~n_{(0)}~.
\end{equation}
Here, $m$ denotes the $m$-th ionization stage with respect to the uncharged atom (reference state). 
Accordingly, $n_{(0)}$ and $n_{(m)}$ are the densities of particles in these states. 
Keeping the total particle density fixed at the bulk value $n_{\textrm{\scriptsize tot}}$ the following conservation laws have been used to solve the system of Saha equations
\begin{equation}
n_{\textrm{\scriptsize tot}}=\sum_{m=0}^Z n_{(m)} \textrm{~~and~~} n_e=\sum_{m=0}^Z m\cdot n_{(m)}\label{ne-equation}~.
\end{equation}
Here, we limit ourselves to a maximum number of ionization stages sufficient for temperatures we are interested in ($T \le 250$ eV) , 
i.e. the sum doesn't run until 18 but is considered up to the Lithium like ion ($m=15$).

Let us now consider the electronic partition function $\sigma_e^{in}$. 
Since the electrons do not have any internal degrees of freedom, we split the partition function into the ideal part represented by the statistical weight $g_e=2$ and a Boltzmann factor containing the interaction energy $\Delta_e$,
\begin{equation}
\sigma_e^{in}=g_e~e^{-\Delta_e/k_BT}~.
\end{equation}
The electron interaction energy $\Delta_e$ is treated as a rigid,  momentum independent quasi-particle shift of the free particle energies and is divided into an electron-ion interaction term and an electron-electron interaction term. According to \cite{CPP_PL} the total internal partition function of the non-degenerate free electrons reads
\begin{eqnarray}
 \sigma_{e}^{in}(m) &=&2^{m} \exp\left[-\frac{\Delta_{e}^{\textrm{MW}}}{k_{B}T} \right]^{m}\\
 &&\times\exp \left[\frac{1}{k_{B}T} \frac{e^{2}}{(4\pi)^{2/3}\varepsilon_{0}} \left(\frac{n_{e}}{3}\right)^{1/3} \sum_{x=1}^m x^{2/3} \right].\nonumber
\end{eqnarray}
The first factor is given by the statistical weight, the second factor is given by the electron-electron interaction and the third factor originates from the electron-ion-interaction. 
For the contribution of the electron-electron interaction we use the Montroll-Ward approximation \cite{saha}
\begin{eqnarray}
\Delta_{e}^{\textrm{MW}}&=&-\frac{e^2}{2}~\sqrt{\frac{n_ee^2}{\varepsilon_0k_BT}}\nonumber\\
&&+\frac{\sqrt{2\pi^{2}}n_{e}\lambda_{e}e^{4}}{8(k_BT)^{2}}-\frac{n_{e}\lambda_{e}^{3}}{8\sqrt{2}}
+\frac{n_{e}\lambda_{e}^{2}e^{2}}{4k_BT}.
\end{eqnarray}

Taking a closer look at the the internal partition functions of the ions $\sigma_{(m)}^{in}$ we have to consider internal degrees of freedom, i.e. differently excited bound states. 
The screening due to the plasma environment leads to a shift of the ionization energies till the bound states vanish at the Mott densities (pressure ionization)\cite{Lin,quantum}. To avoid discontinuities due to pressure ionization, we apply a Planck-Larkin renormalization to the internal partition function \cite{CPP_PL}. It than reads for the $m$-th ionization stage as follows
\begin{eqnarray} 
\sigma_{(m)}^{in}&=&\textrm{exp}\left[-\frac{E_0^{(m+1)}}{k_BT}\right]\\
&&\times~\sum_i\left(\textrm{exp}\left[-\frac{\Delta E_i^{(m)}}{k_BT}\right]-1+\frac{\Delta E_i^{(m)}}{k_BT}\right).\label{PL}\nonumber
\end{eqnarray}
To obtain this equation we separate the energies $E_i^{(m)}$ of possible bound states of the $m$-th ion 
into the ground state energy of the next ionization stage $E_0^{(m+1)}$ and possible excitations of the outermost valence electron 
$\Delta E_i^{(m)}$,
\begin{equation}
E_i^{(m)} =E_0^{(m+1)} + \Delta E_i^{(m)}~. 
\end{equation}
Within this approximation a Mott transition would refer to a state where $\Delta E_i^{(m)}\rightarrow 0$ due to plasma polarization. 

\begin{figure}
 {\includegraphics[width=0.4\textwidth]{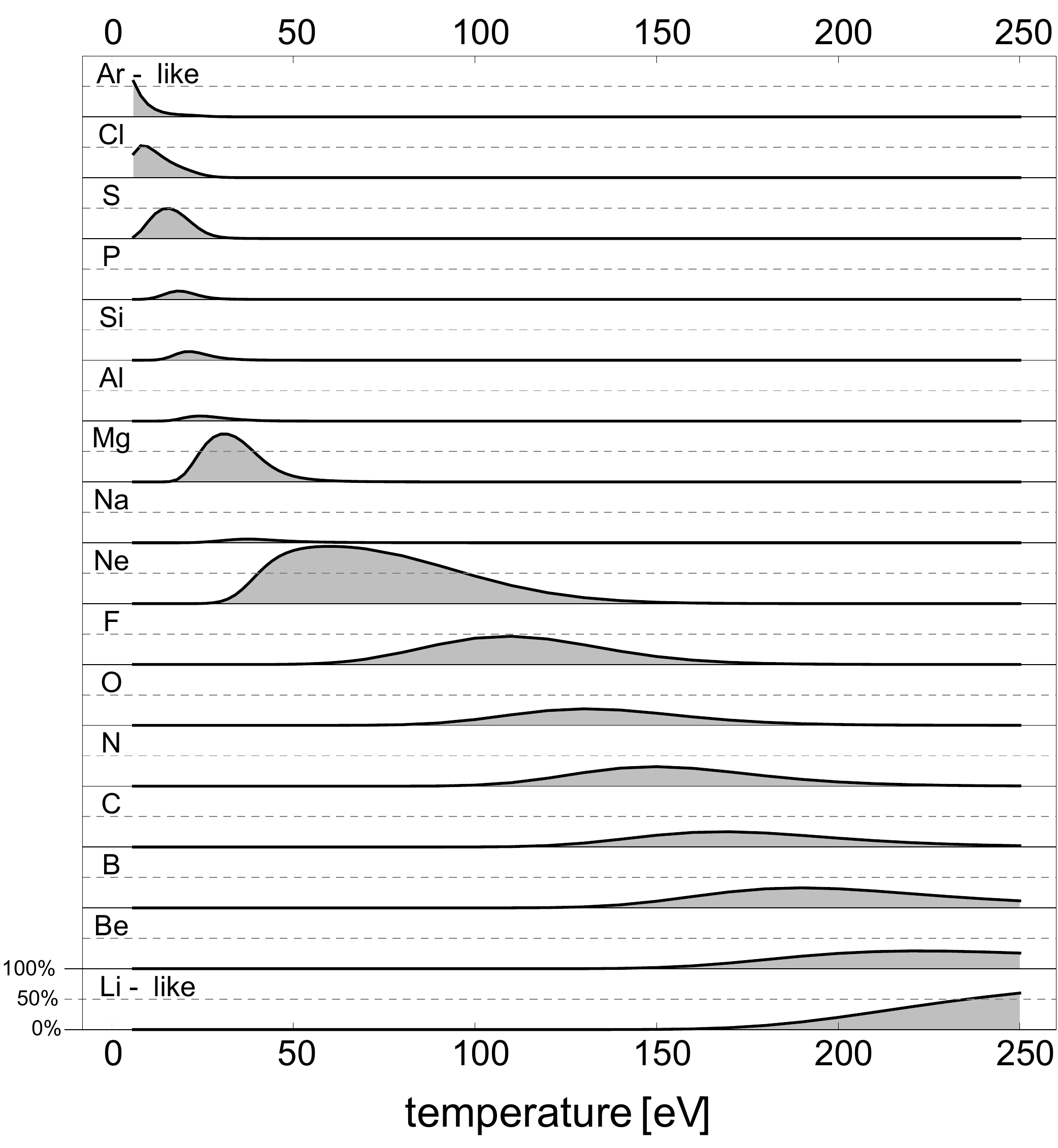}}
	\caption{Plasma composition depending on bulk temperature: percental abundance of different Ar ions.}\label{compo}
  \end{figure}

\begin{figure}
{\includegraphics[width=0.4\textwidth]{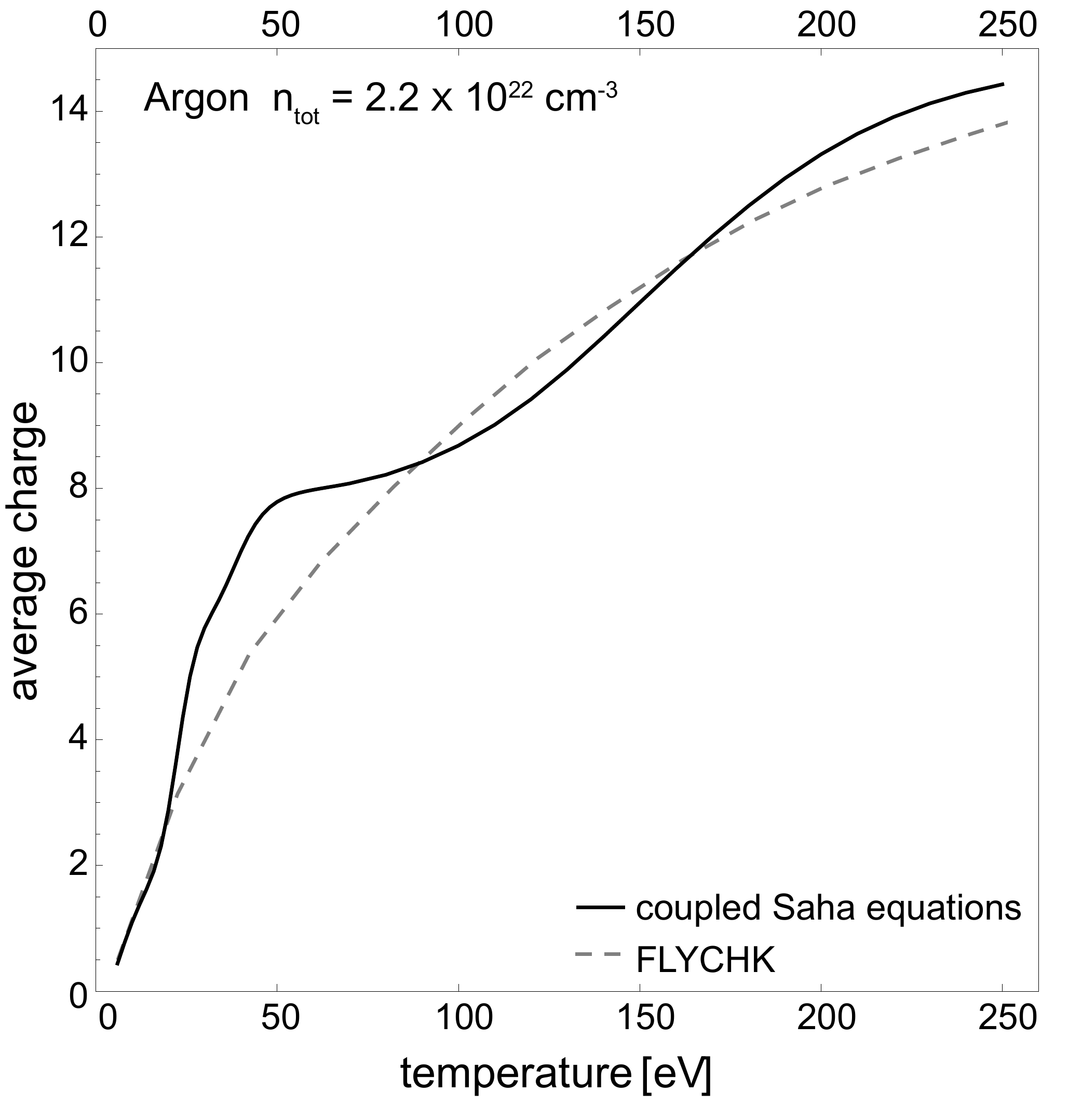}}
  \caption{Average charge state of the plasma depending on bulk temperature. (solid) this work, (dashed) from Neumayer \textit{et al.} obtained by FLYCHK\cite{Ar}.}\label{charge}
  \end{figure}

Keeping the total density fixed at the bulk value of $n_{\textrm{\scriptsize tot}} = 2.2 \times 10^{22} $ cm$^{-3}$ we calculate 
the abundance of different ion species (figure (\ref{compo})) and the corresponding average charge state  (solid line in figure (\ref{charge})) depending on the plasma 
temperature. We compare our results for the charge state distribution with results of Neumayer \textit{et al.} 
obtained from the FLYCHK code \cite{Ar} (dashed line in figure (\ref{charge})). 
We see a sharp rise of the charge state and a rapid sequence of ionization stages are passed through for temperatures below 50 eV. 
This is the regime of M-shell ionization, binding energies are rather low and pressure ionization plays an important role. Comparing with FLYCHK, at the same temperature our calculations give not only a higher charge state but also larger energy shifts of the emission energies in this regime. For temperatures above 50 eV, where the L-shell ionization regime sets in, our results swing back to the FLYCHK results and oscillate around the graph. 

The differences in figure (\ref{charge}) arise from the different treatment of the variety of bound states and pressure ionization within the approaches. 
As there are several hundred transitions closely distributed for lower charge states usually a simplification is applied for better handling. 
FLYCHK uses super-configuration transition arrays (effective bound states) weighted by respective transition rates rather than the complete level structure\cite{Ar}. 
We do not average over several levels, but take all low lying excited states into account. 
Further, in the FLYCHK code continuum lowering is treated within a Stewart-Pyatt like model to implement medium corrections to the ionization energies. 
The model interpolates between the two limiting cases: Debye screening for dilute and hot plasma on the one hand and screening within an ion sphere model with constant free electron density for dense and cool plasmas an the other hand\cite{FLYCHKManual}. If bound states merge into the continuum, the calculations get discontinuous since these states suddenly disappear 
from the implemented rate equations. Using effective bound states reduces these events. Within our quantum statistical approach, there is no such discontinuity thanks to the use of the Planck-Larkin renormalization of bound states \cite{CPP_PL}. As shown above, we also use an ion sphere model to calculate the continuum lowering. However, the free electron density within the sphere is not constant but radially dependent. 
All in all, we observe several Mott transitions in solid density argon at temperatures below 50 eV which lead to a higher degree of ionization than proposed by FLYCHK. Looking at figure (\ref{compo}) magnesium and neon like states seem to be more stable, which is due to their closed shell structure. Hence, the slope of the charge state in figure (\ref{charge}) flattens for rising plasma temperatures. As the ion charge rises pressure ionization becomes less important and our results converge to the FLYCHK values. The oscillation around the values is in accordance with the variing shell structure of the ions. In agreement with the expectations, less than half filled shells seem to be the least stable states and closed shells show a high stability against temperature increase. In contrast to our detailed slope, the FLYCHK code creates a rather generic output.  

\section{Synthetic spectra and temperature distribution}

To construct synthetic spectra from the shifted emission energies every line as well as its fine structure components 
are assigned a Lorentz profile with natural line width $\gamma$ and maximum intensity $I_\textrm{max}$,
\begin{equation}
I(E)=\frac{I_\textrm{max}~(\gamma/2)^2}{(E-E_0)^2+(\gamma/2)^2}~.
\end{equation}
The central line position $E_0$ is given by the plasma shifted line positions. The maximum intensity is mainly given by the emitter abundance in the plasma. 
We determine all intensities relative to a reference state, 
\begin{equation}
 \frac{I_\textrm{max}}{I_\textrm{ref}}\simeq \frac{n}{n_\textrm{ref}} \left( \frac{E_{0}}{E_{0}^\textrm{ref}}\right)^{4} .\label{intensity}
\end{equation} 
For the electron transitions considered here, we assume nearly constant dipole matrix elements $\left|\frac{1}{\Omega_0}\int d^3r ~\varphi_\textrm{\scriptsize in}^* ~\vec{r}~\varphi_\textrm{\scriptsize fin}^*\right|^2$ with respect to excitation and ionization of outer shells since the one particle wave functions $\varphi_\textrm{\scriptsize in}$ (initial state, 2p)  and $\varphi_\textrm{\scriptsize fin}$ (final state, 1s) change only very slightly \cite{Sengebusch09}. Hence, the Einstein coefficients for spontanious emisssion are simply $\propto E_{0}^3$ and equation (\ref{intensity}) can easily be derived.

\begin{figure}
{\includegraphics[width=0.4\textwidth]{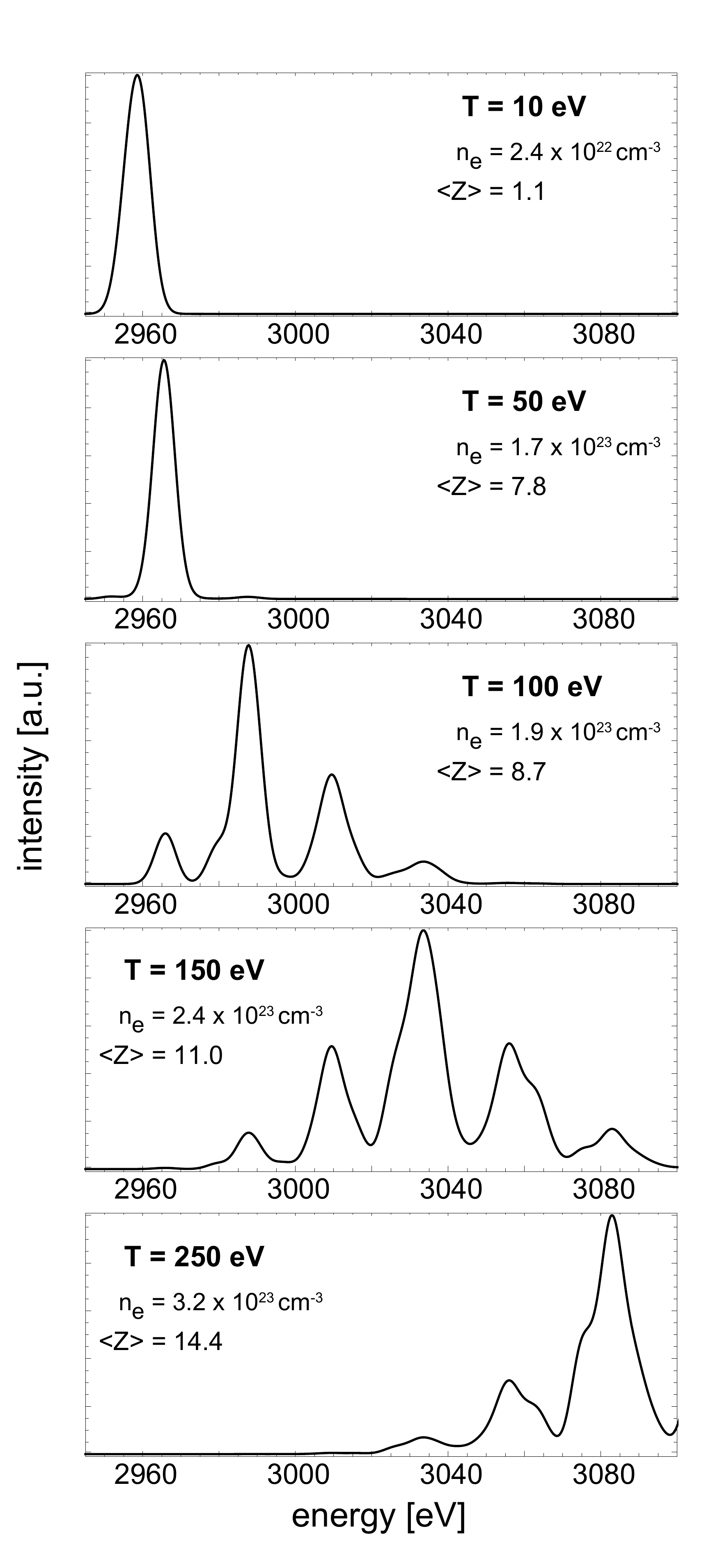}}
  \caption{Selected normalized synthetic spectra for different plasma temperatures. 
  The total density is fixed at the bulk value, the free electron density corresponds to the plasma composition.}\label{Spektren}
  \end{figure}

The Lorentzians are summed up and convoluted with a Gaussian profile of width $\Gamma$ to take into account instrumental broadening 
of the measurements. According to Neumayer \textit{et al.} \cite{Ar} we varied the broadening parameters within the range of some eV 
and found the most suitable combination in a universal natural linewidth of $\gamma=1$ eV and a Gaussian width of $\Gamma= 2.5$ eV. 
A selection of spectra obtained at different plasma temperatures is shown in figure (\ref{Spektren}). 
Keeping the total density and plasma temperature fixed, the free electron density follows from the composition. 
Shown are some typical spectra: At low temperatures, up to some 10 eV, we are in the regime of M-shell ionization. 
Since the differences of the various emission energies are small, the lines of Ar$^+$ to Ar$^{9+}$ group indistinguishably within one large main peak which 
shifts to higher energies with rising temperatures. With further increase of the temperature, we enter the regime of L-shell ionization. 
The main peak becomes less prominent and several distinguishable peaks appear, which correspond to the emission energies 
of the respective ions. We calculated spectra up 250 eV where the Li-like satellites are dominant. 
Experimentally observed higher emission energies mainly arise from argon ions with only one or two electrons (Ly-$\alpha$, He-$\alpha$,
intercombination line). These emissions stem from the blow-off plasma created by the laser impact.

\begin{figure}
{\includegraphics[width=0.4\textwidth]{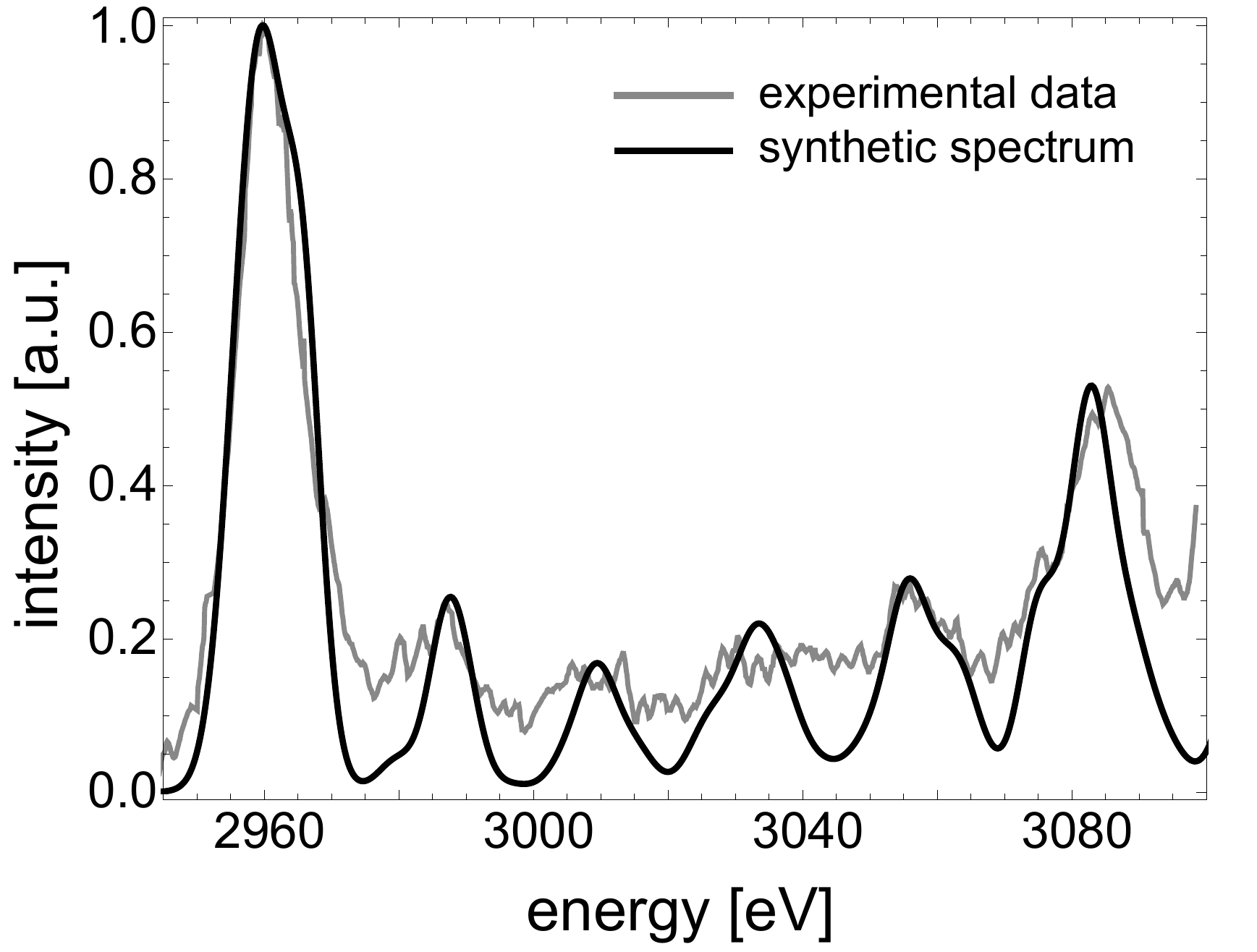}}
  \caption{Best fit of superposition of theoretical spectra to space-time-integrated measurements 
  of Neumayer \textit{et al.} \cite{Ar}.}\label{bestfit}
  \end{figure}

\begin{figure}
{\includegraphics[width=0.4\textwidth]{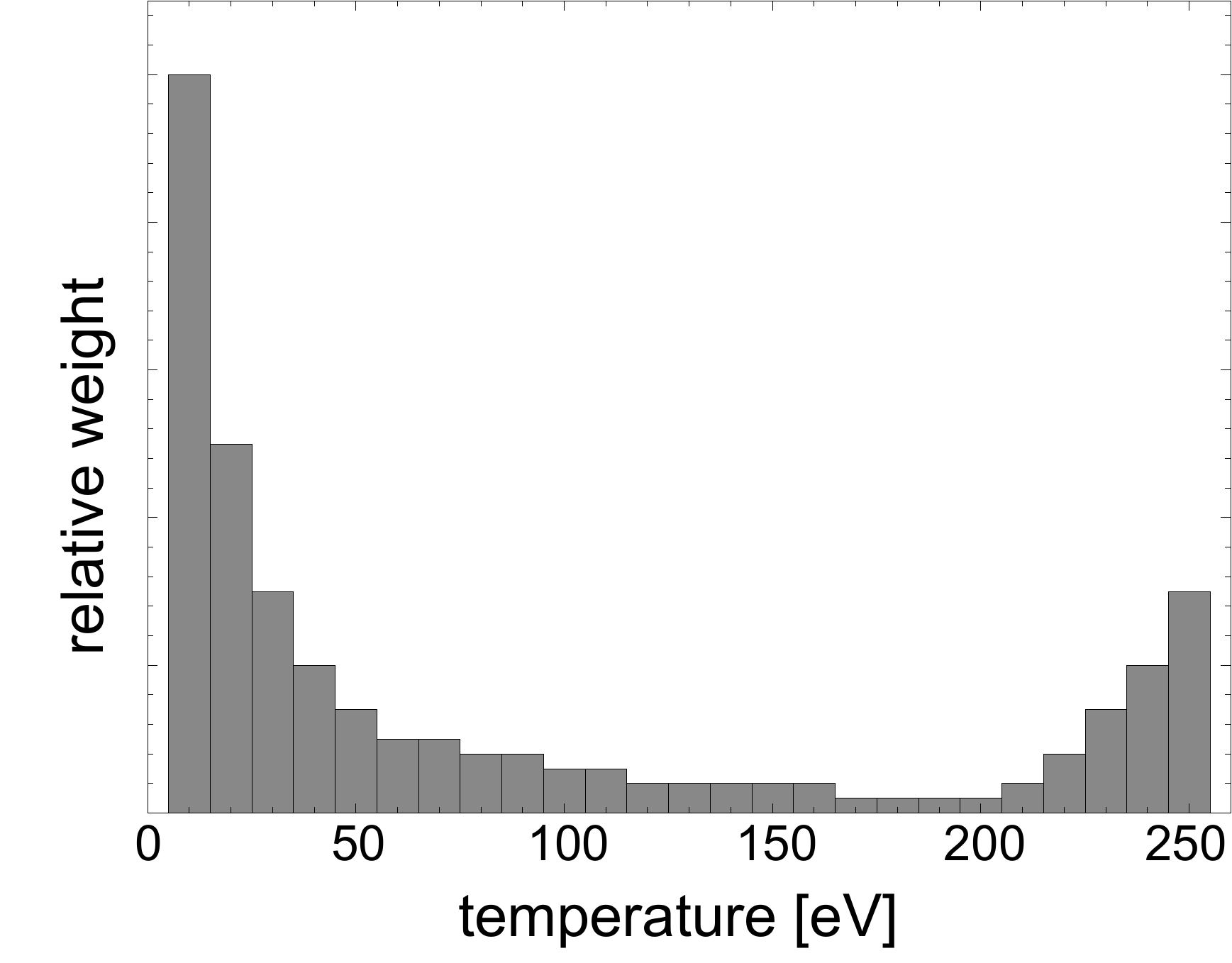}}
  \caption{Temperature distribution which gives the best fit of theoretical to experimental spectra. 
  The relative weights are determined via the method of least squares.}\label{temperaturedistribution}
  \end{figure}

We apply a superposition of our calculated spectra for different temperatures to model the experimental results obtained by Neumayer \textit{et al.} \cite{Ar}. 
To obtain the best fit the weight of the different spectra is determined within a variational approach using the method of least squares. 
Results for the best fit and the corresponding temperature distribution are shown in figure (\ref{bestfit}) and figure (\ref{temperaturedistribution}), respectively. 
The superimposed spectrum shows a good agreement especially for the M-shell satellites. The position and width of the main peak is rather sensitive to the plasma bulk temperature. 
At a first glimpse, the agreement with the L-shell satellites is less satisfying. The peaks are at correct positions but the widths seem to small. We attribute this to our choice of the natural line widths. We use a universal value of $\gamma$ which implies the same lifetime for all considered states. But especially highly ionized and excited states experience shorter lifetimes and hence show a larger broadening. Accordingly, the L-shell emission lines would blur and give a better fit to the energy spectrum above 2980 eV.

\section{Discussion and conclusion}

The temperature distribution, which best fits to the experimentally obtained argon K$_{\alpha}$ spectra, is a two-peaked curve with local maxima at our lowest temperature value of 10 eV and at our highest temperature value of 250 eV and a steady decrease of relative weights for intermediate temperatures. We chose temperature steps of 10 eV to keep the variation manageable. Smaller steps are possible, however the two peaked slope of the temperature distribution would not change significantly. The two peaks can be interpreted as a result of the two differently heated parts of the argon droplet, namely the cold bulk and hot blow off plasma. The blow off plasma is directly created by the laser-plasma interaction. It is rather hot and dilute and is indicated by the H- and Ne-like emission lines above 3100 eV. However, as the experimental data is spatially and temporally integrated, the measurements contain also emissions of not fully heated or already cooled down particles. Due to such emitters we obtain the high temperature tail above 200 eV. 
In fact, this work is more related to the bulk temperature distribution due to electronic heating. The electrons are accelerated in the laser field and flow directedly trough the droplet, mainly heating the area close to the laser focus and the volume behind. As the electrons scatter and the ions tend to equilibrate, the heat is transferred radially to cooler regions of the droplet. The maximum of the temperature distribution at lowest temperatures originates on the one hand from emitters of the outer spacial regions and on the other hand from particles that already cooled down before emission.

There are significant differences between the temperature distribution proposed by Neumayer \textit{et al.} \cite{Ar} and the results shown here: according to their FLYCHK results, they also obtained a two peaked distribution, only our first maximum lies at much lower plasma temperature and our temperature range is about 100 eV broader. These differences are due to the different results of the composition calculations as show in the previous chapter and figure (\ref{charge}). As we obtain higher degrees of ionization at lower temperatures, we see the general shift of emission lines to higher energies (blue shift) already at lower temperatures and hence we obtain a temperature of about 10 to 20 eV instead of 50 eV to fit the main peak of the measured spectrum. Neumayer \textit{et al.} discussed several mechanisms like resistive heating or radiative heating that might help to explain the absence of temperatures below 50 eV. However, our results suggest, that these influences might not be as substantial as the authors assumed. In order to gain more insight into the heating and cooling processes within the plasma, kinetic codes and hydrodynamic simulations are favourable.

The details of the synthetic spectra need further discussion. It turns out, that using the same broadening for all lines slightly overestimates the width for the main peak (Ar$^{+}$ to Ar$^{9+}$) and underrates the width for the higher lying peaks as pointed out above. A more detailed discrimination of the widths would improve the agreement between theory and experimental data. In particular, the theoretically obtained low minima between the peaks will possibly be smeared out, if the applied broadening is dependent on the emitter configuration as well as on the plasma parameters. A quantum statistical approach to the line width would be desirable, which is well elaborated for H-like systems \cite{broadening} but needs further development for mid-$Z$ materials.

The inclusion of plasma effects, especially shifts and the merging of bound states with the continuum, is important to discuss the composition of the plasma and the density of free electrons. 
We have shown that such effects have significant influence on the inferred temperature distribution. In general we have shown that K$_{\alpha}$ spectroscopy is an interesting means for plasma diagnostics and allows an estimate of the temperature profile in warm dense matter.

\ack
We would like to thank P. Neumayer for fruitful discussions and the experimental data. 
This work has been supported by the German Research Foundation through the Collaborative Research Center 652 - Strong 
correlations and collective effects in radiation fields: Coulomb systems, clusters and particles.

\section*{References}


\end{document}